%% file: ms.tex
\documentclass[12pt, preprint]{aastex}

\newcommand{\farc}{\hbox{$.\!\!^{\prime\prime}$}} 
\begin{document}


\title{The 2175~\AA~dust feature in a Gamma Ray Burst afterglow at redshift 2.45}

\author{T. Kr\"{u}hler\altaffilmark{1,2}, A. K\"{u}pc\"{u} Yolda\c{s}\altaffilmark{2}, J. Greiner\altaffilmark{2}, C. Clemens\altaffilmark{2}, S. McBreen\altaffilmark{2}, N. Primak\altaffilmark{2}, S. Savaglio\altaffilmark{2}, A. Yolda\c{s}\altaffilmark{2}, G. P. Szokoly\altaffilmark{2,3}}
\and
\author{S. Klose\altaffilmark{4}}

\altaffiltext{1}{Universe Cluster, Technische Universit\"{a}t M\"{u}nchen, Boltzmannstra\ss e
2, D-85748, Garching, Germany; kruehler@mpe.mpg.de}
\altaffiltext{2}{Max-Planck-Institut f\"{u}r extraterrestrische Physik, Giessenbachstra\ss e, D-85748 Garching, Germany}
\altaffiltext{3}{Present Address: Institute of Physics, E\"{o}tv\"{o}s University,
P\'{a}zm\'{a}ny P. s. 1/A, 1117 Budapest, Hungary}
\altaffiltext{4}{Th\"{u}ringer Landessternwarte Tautenburg, Sternwarte 5, D-07778 Tautenburg, Germany}

\begin{abstract}
We present optical and near-infrared photometry of the afterglow of the long Gamma-Ray Burst GRB~070802 at redshift 2.45 obtained with the ESO/MPI 2.2~m telescope equipped with the multi-channel imager GROND. Follow-up observations in $g\arcmin r\arcmin i\arcmin z\arcmin$~and $JHK_S$~bands started $\sim$17~min and extended up to 28~h post burst. We find an increase in brightness of the afterglow at early times, which can be explained by the superposition of reverse and forward shock (FS) emission or the onset of the afterglow FS. Additionally, we detect a strong broad-band absorption feature in the $i\arcmin$~band, which we interpret as extinction from the redshifted 2175~\AA~bump in the GRB host galaxy. This is one of the first and clearest detections of the 2175~\AA~feature at high redshift. It is strong evidence for a carbon rich environment, indicating that Milky Way or Large Magellanic Cloud like dust was already formed in substantial amounts in a galaxy at z=2.45.
  
\end{abstract}


\section{Introduction}

Gamma-Ray Bursts (GRBs) are intense extragalactic flashes of $\gamma$-rays with durations between several tenths to hundreds of seconds \citep[e.g.][]{fis94, mes06}. They cluster in two different categories of duration and spectral hardness: short hard bursts versus long soft bursts \citep{kou93}. GRBs are followed by longer lasting afterglows in radio to X-rays \citep[e.g.][]{kat94, mes97, zha07}. The enormous energy release in $\gamma$-rays is unaffected by dust or gas absorption, so GRBs are detectable out to very high redshifts \citep[e.g.][]{lam00, kaw06}. These characteristics make GRBs potential tools to constrain the history of star formation \citep{bro02} and the chemical evolution in the universe \citep{sav06, fyn06, ber06}.

The GRB-SN connection \citep[e.g.][]{zeh04, woo06} strongly supports the association of long GRBs with the core-collapse of very massive stars, and therefore with regions of high-mass star formation in the host galaxies \citep{pac98}. High-mass star forming regions show signatures of significant dust and gas absorption in ultra-violet and optical spectra of bright GRB afterglows \citep{sav06, ber06, fyn07}. In fact, a large number of bursts do not show optical afterglows despite rapid and deep optical follow-up observations \citep[e.g.][]{str04, rol05}. Natural explanations for 'optically dark bursts' \citep{gro98} include dust extinction in the host galaxy \citep[e.g.][]{fyn01, klo03} and high redshift \citep[e.g.][]{gro98}, both efficiently suppressing any flux in the observed optical bands \citep{rom06}. Several bursts with moderate dust absorption have been detected \citep[e.g.][]{kan06}, which may represent only the lower end of the host extinction distribution \citep{sch07}. The detection of significant reddening in a host is strongly instrumentally biased due to the lack of rapid near-infrared follow-up observations with large aperture telescopes. In particular, one of the biggest shortcomings of all multi-color monitoring of afterglows so far was the lack of a simultaneous coverage of the optical to near-infrared (NIR) bands. 

The Gamma-Ray Burst Optical and Near Infrared Detector (GROND) is a 7-channel imager primarily designed for fast follow-up observations of GRB afterglows \citep{gre07a, gre08}. It enables the detection and identification of GRB afterglows in a broad wavelength range (400~nm - 2310~nm). Due to the use of dichroic beamsplitters it is capable of simultaneous imaging in seven bands, $g\arcmin r\arcmin i\arcmin z\arcmin$ (similar to the Sloan system) and $JHK_S$. GROND is mounted on the 2.2~m ESO/MPI telescope on LaSilla/Chile since April 2007. The instrument is operated robotically and capable of monitoring the light curve of the transient starting from a minimum of a few minutes after burst alert. The field of view of the instrument is 10$\arcmin$~$\times$~10$\arcmin$ in the NIR and 5.4$\arcmin$~$\times$~5.4$\arcmin$ for the optical bands \citep{gre08}. The fast response, medium sized telescope aperture, NIR capabilities and unique optical design makes GROND an ideal tool for follow-up observations of GRBs.

Here we report on first GROND follow-up observations of GRB~070802 (section \ref{obs}) and derive constraints on the GRB ejecta and its circumburst properties (section \ref{ana}).

\section{Observations}
\label{obs}
\subsection{Swift observations}

The BAT (Burst Alert Telescope) instrument \citep{bar05} onboard the \textit{Swift} satellite \citep{geh04} triggered on the long-soft GRB~070802 at $T_0$=07:07:25~UTC and immediately slewed to the burst \citep{bar07}. 
The BAT light curve shows a single peak starting at T$_0$+5 s and ending at T$_0$+50s. There is evidence at the 3$\sigma$ level for a precursor at T$_0$-150 s \citep{cum07}. The T$_{90}$ for GRB 070802 is 16.1$\pm$1.0~s and the fluence in the 15~keV to 150~keV band is 2.8$\pm$0.5 $\times$ 10$^{-7}$ erg cm$^{-2}$ \citep{cum07}. 

The X-Ray Telescope (XRT, \citealp{bur05a}) began follow-up observations of the burst field 138~s after the trigger and detected an uncatalogued fading X-ray source at a position of Ra(J2000.0)=02$^h$~27$^{min}$~35$^s$.76, Decl(J2000.0)=-55$\arcdeg$~31$\arcmin$~38\farc4 with a refined 90\% coincidence error of 2\farc1 \citep{man07}. The XRT light curve decays with a slope of $\alpha\sim-2$ until T$_0$+500~s and then remains flat until $\sim$4~ks. Afterwards the light curve decays with a power law index of $\sim-1$ until it fades below the XRT sensitivity limits \citep{man07}. No bright flares are detected in the XRT light curve.

The third instrument onboard \textit{Swift}, the Ultra-Violet Optical Telescope (UVOT, \citealp{rom05}) started observations at T$_0$+100~s and did not find any transient sources inside the XRT error circle down to 19.5 (u filter), 21.3 (b filter) and 21.2 (uvw2 filter) magnitudes \citep{imm07}. 

\subsection{GROND optical and near infrared observations}

\begin{figure}
\epsscale{1.0}
\plotone{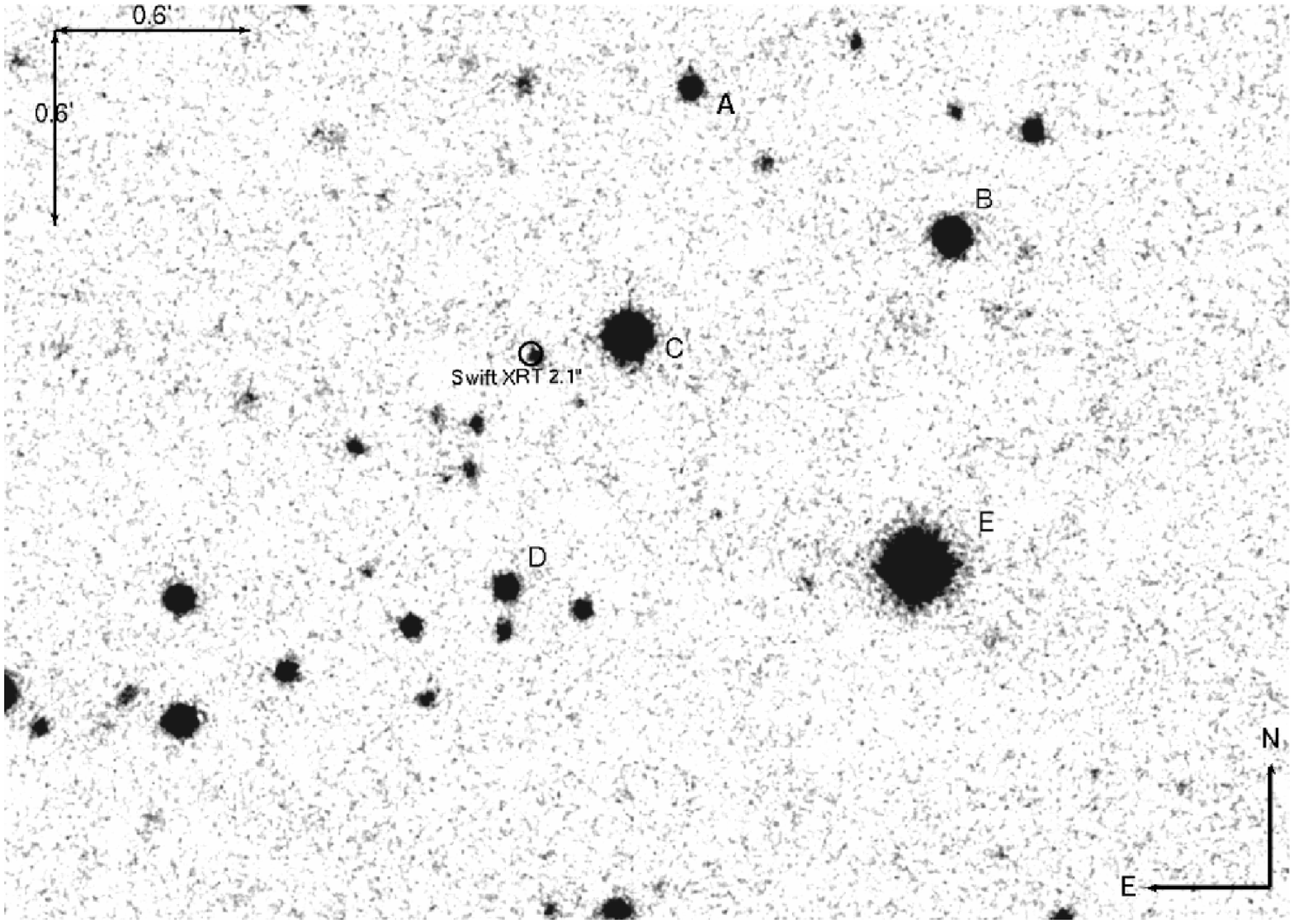}
\caption{NIR afterglow of GRB~070802 in the $J$ band at t=3.4$\pm$0.3~ks post burst, including the Swift XRT error circle. Only a $\sim$3\arcmin~$\times$~2.5\arcmin~segment of the original 10\arcmin~$\times$~10\arcmin~image is shown. The GROND NIR images have a pixel scale of 0\farc6/px each. The image is a combination of 48 stacked 10~s exposures and also shows the secondary standards used for calibration, marked as A, B, C, D and E.}
\label{pic1}
\end{figure}

GROND responded to the  \textit{Swift} GRB alert and initiated automated observations on the 2$^{nd}$ of August 2007 at 07:24:09 UTC, starting 7~min 16~s after the \textit{Swift} trigger and 16~min 44~s after the onset of the burst. A predefined sequence of exposures with successively increasing exposure times were executed, acquiring images in all seven photometric bands simultaneously. The observations continued for two nights, after which the afterglow had faded below the GROND sensitivity limits. A variable point source in the NIR bands \citep{gre07} inside the \textit{Swift} XRT error circle was identified by the GROND data reduction pipeline (A. Yolda\c{s} et al. 2008, in preparation). The transient is shown in Fig.~\ref{pic1} and its absolute position is calculated to be Ra(J2000.0)=02$^h$~27$^m$~35$^s$.68, Decl(2000.0)=-55$\arcdeg$~31$\arcmin$~38\farc9 with an uncertainty of 0\farc3 compared to 2MASS reference field stars. The afterglow was also observed and detected by the Magellan telescope at Las Campanas Observatory (LCO, \citealp{ber07}) and the Very Large Telescope (VLT), the latter yielding a spectroscopic redshift of z=2.45 \citep{pro07}. 

In total 2036~NIR images with an integration time of 10~s each, and 56~CCD optical frames were obtained with GROND. The CCD integration times scaled with the brightness of the transient from 45~s at early times to 10~min when the source had faded. Sky conditions were clear with a mean seeing around 1\farc2. All GROND data were obtained at airmasses between 1.35 and 1.12. 

Optical and near-infrared image reduction and photometry was performed using standard IRAF tasks \citep{tod93}. A general model for the point-spread function (PSF) of each image was constructed using bright field stars and fitted to the afterglow. Additionally, aperture photometry was carried out and the results were consistent with the reported PSF photometry. 
Photometric calibration was performed relative to secondary standards in the GRB field. During photometric conditions, two spectrophotometric standard stars, SA114-750 and SA114-656, both primary Sloan standards \citep{smi02}, were observed with GROND. Observations of the GRB field followed within 4 minutes. The magnitudes of SA114-750 and SA114-656 were transformed to the GROND filter system using their spectra and the GROND filter curves \citep{gre08}. The obtained zeropoints were corrected for atmospheric extinction differences and used to calibrate stars in the GRB field, shown in Fig.~\ref{pic1}. The apparent magnitudes of the afterglow were measured with respect to the secondary standards reported in Table \ref{tabSecStan}.

Vega magnitudes have been transformed to the AB system using transformation factors for the GROND filter system as $\delta g\arcmin$=0.01~mag, $\delta r\arcmin$=0.15~mag, $\delta r\arcmin$=0.39~mag, $\delta z\arcmin$=0.52~mag, $\delta J$=0.91~mag, $\delta H$=1.38~mag and $\delta K_S$=1.80~mag. All reported afterglow magnitudes are corrected for Galactic foreground reddening ($E_{B-V}$=0.026 mag, \citealp{sch98}). Assuming $R_{V}$=3.1 for the Milky Way, this leads to $A_K$=0.01, $A_H$=0.02, $A_J$=0.03, $A_z$=0.04, $A_i$=0.05, $A_r$=0.07 and $A_g$=0.10 for the GROND filter bands.

The afterglow is detected in all seven GROND bands, however it was too dim to construct a light curve with reasonable time resolution in the filter bands $g\arcmin$ and $i\arcmin$. The light curves obtained in the $r\arcmin z\arcmin JHK_S$ bands are presented in Fig.~\ref{picLight}. The light curve behaviour is dominated by an early rise in brightness, after which it declines with a bump superimposed onto the overall decay. The observed variations occur in all five GROND bands, and the generic light curve shape is achromatic within the measurement uncertainties.

\section{Analysis}
\label{ana}
\subsection{The early light curve of GRB~070802}

According to the fireball model \citep[e.g.][]{wij97, sar99, pir05, mes06}, GRBs produce their prompt emission in $\gamma$-rays from internal shocks of an ultra-relativistic outflow from a compact source and long-wavelength afterglows from the interaction of the ejecta with the circumburst medium. After the prompt internal shock phase, the optical afterglow light curve is composed of a superposition of two different emission components. The reverse shock (RS) propagating into the ejecta and the forward shock (FS) travelling into the surrounding medium \citep{zha03}. Rapid optical observations of the early transition phase between prompt and afterglow emission can constrain the nature of the outflow \citep{nak04}. While baryonic ejecta are expected to produce an optical flash, that can be associated with a RS, a Poynting flux dominated afterglow should preferentially show the FS emission. The hydrodynamical calculations from the fireball model have succeeded in describing the generic afterglow of GRBs from several minutes to days post burst. However, the majority of bursts do not show bright optical flashes and apparently lack a strong RS emission component \citep{rom06}. This fact provides observational support that the strength of the optical emission from the RS may be weaker than previously calculated \citep{bel05}. 

A baryonic shell expanding into a homogeneous medium is fully described by the shell isotropic equivalent energy E, its width $\Delta$, the initial Lorentz factor $\Gamma_0$ and the ISM density \textit{n} \citep{nak04}. In the thick shell case, the initial Lorentz factor is large, and the swept up circumburst medium decelerates the ejecta efficiently \citep{kob00}. Here, the RS becomes relativistic in the rest frame of the unshocked ejecta. The critical Lorentz factor $\Gamma_c$ discriminating between relativistic thick ($\Gamma_0 >\Gamma_c$) and Newtonian thin shell case ($\Gamma_0 <\Gamma_c$) is given by \citep{zha03}:
\begin{equation}
\Gamma_c \cong 125 E_{52}^{1/8}n^{-1/8}T_2^{-3/8}\left ( \frac{1+z}{2}\right )^{3/8}
\label{eq1}
\end{equation}
where $E_{52}$ is the isotropic energy equivalent in units of $10^{52}$~erg and T$_2$ is the burst duration in units of 100~s. For the thin shell case the Lorentz factor at the crossing time $t_x$ of reverse and forward shock is $\Gamma_x=\Gamma_0$. By measuring the peak of the RS, one can directly derive $\Gamma_0$ for the thin shell case:
\begin{equation}
\Gamma_x = \Gamma_c (T/t_x)^{3/8}
\label{eq2}
\end{equation}
with only a weak dependence on $E_{52}$/\textit{n}. 

The light curve shown in Fig.~\ref{picLight} displays a behaviour consistent with the upper theoretical predictions. It rises achromatically at early times and reaches a first peak at $\sim$2~ks. Afterwards the light curve declines with an power law index of $\sim$-2 until $\sim$3~ks. Between 3~ks and 10~ks post burst our data are not conclusive. There may be a plateau phase, although a power law decline with the late time index cannot be ruled out. At late epochs the afterglow follows the canonical power law decay. The data are compatible with a constant power law spectrum at all times, although there is an indication for chromatic changes at peak brightness. There is no evidence for a jet break in the GROND data out to 90~ks post burst.

We used two different approaches to analyze the light curve quantitatively. Firstly, a physical model combining the forward and reverse shock component (RS-FS model), and secondly a forward shock model alone (FS model).

Within the framework of the combined RS-FS model, the first peak can be interpreted as the peak of the reverse shock, whereas the possible rebrightening phase around 4~ks is related to the FS peak. 
A complete parametrization of the RS is given by a broken power law \citep{beu99, nak04}: 
\begin{equation}
F_{\nu}^{r}(t) = F_0^r\left[\left( \frac{t}{t_1}\right)^{-s^r\alpha_1^r} +\left( \frac{t}{t_1}\right)^{-s^r\alpha_2^r} \right]^{-1/s^r}
\end{equation}
with power law indices of rise ($\alpha_1^r\geq $0) and decline ($\alpha_2^r\sim$ 2), a peak time $t_{peak}=t_1(\alpha_1^r/-\alpha_2^r)^{1/(s^r(\alpha_1^r-\alpha_2^r))}$, normalization $F_0^r$, and the sharpness of the break $s^r$. 

The FS parametrization used for the light curve analysis is similar to the RS:
\begin{equation}
F_{\nu}^{f}(t) = F_0^f \left[\left( \frac{t}{t_2}\right)^{-s^f\alpha_1^f}+\left( \frac{t}{t_2}\right)^{-s^f\alpha_2^f} \right]^{-1/s^f}
\end{equation}
According to theoretical calculations \citep{sar98}, a power law index of the FS with $\alpha_1^f \approx 1/2$ is expected for the forward shock rise, followed by a shallow decline $\alpha_2^f\approx 3/4 - 3/4 p$. Here \textit{p} is the power law index of the energy distribution of the shocked electrons \citep{sar98, van00}. The power law indices for both forward and reverse rise and decline as well as the peak time have been fitted using the complete multi-color data set. The flux normalizations result from single band fits. The result of the combined fit of the RS and FS are consistent with the expected power law decline of the RS with a power law index $\sim-$2 (Tables \ref{tab1} and \ref{tab2}).

To constrain the RS/FS fit, the sharpness parameter of the power law transition ($s^r$) and the time of the putative FS ($t_2$) had to be fixed to 2.5 and 4.5~ks, respectively. However, this affects mainly the late time properties of the afterglow, whereas our analysis is concentrated on the early epoch dominated by the rise of the light curve. Specifically, the peak of the light curve is unaffected by the fixed parameters.

We also fit the data with the FS broken power law alone. In this case the early rise of the afterglow is interpreted as the onset of the afterglow. The single FS component fit is consistent with a power law rise index $\sim$3, as expected for an ISM profile and results in fit parameters as reported in Tables \ref{tab1} and \ref{tab2}. This simple model alone cannot explain the steep decay of the initial peak. A further emission component at $\sim$2.5~ks superimposed to the overall power law rise and decay is required to explain the observed light curve features. Rebrightening episodes and variabilities in the optical afterglow light curve have been observed in a number of previous burst, e.g. GRB~021004 \citep[e.g.][]{fox03}, GRB~030329 \citep[e.g.][]{lip04}, GRB~050502A \citep{gui05}, GRB~061126 \citep{per08} and GRB~070125 \citep{upd08}. Possible explanations include inhomogeneities in the density profile of the circumburst medium \citep[e.g.][]{wan00, laz02}, in the angular distribution of the outflow (i.e. the patchy shell model, \citealt{kum00}) or late energy injection by refreshed shocks \citep[e.g.][]{ree98}. \citet{nak07} find that sharp rebrightenings in the optical light curve are very unlikely to be caused by density jumps in general and favour the refreshed shock (GRB~030329) or patchy shell model (GRB~021004). Points in the afterglow light curve of GRB~070802 which are attributed to the superimposed component at $\sim$2.5~ks have been excluded from the generic light curve fit of the exclusive FS model.

The peak times of both parametrizations can then be used to constrain the initial conditions in the ejecta. Due to the lack of sufficient time resolution between 3~ks and 10~ks, the two models cannot be clearly discriminated. 

Within the interpretation of the early brightening of the afterglow of GRB~070802 as the RS, a wind-shaped circumburst medium and Poynting flux dominated models are implicitly ruled out. In both cases a RS would not cause a $t^{-2}$ decay and it is very unlikely to be imitated by other phenomena \citep{nak04}. 

Although the profile of the light curve of the optical/NIR afterglow shows evidence for a reverse shock in an ISM circumburst medium, there is still a problem with its timing relative to the burst. The putative RS peak at $\sim$2~ks is highly delayed with respect to the duration of the burst (16.4 $\pm$ 1.0~s), which is possible for a very thin shell only. As a consequence, the initial Lorentz factor of the outflow, which is estimated to be $\Gamma_0 \approx 40 (E_{52}/n)^{1/8}$, is quite small compared to the expected distribution of the Lorentz factors above 100 \citep{pir00} and previous bursts \citep{mol07, pee07, fer08}. A possible solution within the context of the RS scenario would be an extremely low density environment.

Alternatively, the initial increase in brightness might be related to the onset of the forward shock of the afterglow itself \citep{pan00}, as suggested for GRB~060418 and GRB~060607A \citep{mol07}. In this case, a RS component might be hidden under the dominating FS emission or occured even before the GROND observations. A possible explanation for the superimposed component occurring at the light curve peak would then be e.g. multiple energy injections by refreshed shocks \citep[e.g.][]{uga05} or dense clumps in the circumburst medium \citep[e.g.][]{gui05}. Flares and rebrightenings are frequently observed in X-ray afterglow light curves \citep{bri06,zha06}. The fact that the length of the flare ($\sim$1200~s) is compatible with the start time of the flare ($\sim$1800~s post burst) supports the refreshed shock scenario. 

Using \citet{sar99}, \citet{pan00} and \citet{mol07}, we estimate $\Gamma_0$ for the ISM case to $\Gamma_0 \approx 160 \left (\frac{E_{53}}{\eta_{0.2}n}\right)^{1/8}$, and for the wind shaped case to $\Gamma_0 \approx 80 \left (\frac{E_{53}}{\eta_{0.2}A^{*}}\right)^{1/4}$, both more in line with previous bursts and theoretical predictions.  Here $E_{53}$ is the isotropic-equivalent energy released in $\gamma$-rays in $10^{53}$~erg, $\eta_{0.2}$ the 0.2 normalized radiative transfer efficiency, \textit{n} the ISM density in cm$^{-3}$ and $A^{*}$ the normalized wind density. The steep rise of the light curve with a power law index of 3.56 $\pm$ 0.36 favours the homogeneous circumburst environment \citep{pan00}.

\subsection{The Spectral Energy Distribution}
\subsubsection{The intrinsic extinction in previous GRB host galaxies}

The GRB/SN connection hints strongly at the progenitor of GRBs, which are supposedly very massive, fast rotating Wolf-Rayet stars \citep[e.g.][]{woo06}. In this collapsar model  \citep{mac99} one would expect long GRBs preferentially in regions with a high star formation rate (SFR) with significant amount of dust and gas in the host along the GRB sight line. However, most previous bursts show only a moderate or low reddening \citep{gal01, kan06, kan08}, contrary to the observed high column densities of heavy elements and strong depletion of refractory elements \citep{sav04} .

The main feature discriminating between the extinction curves in the Large Magellanic Cloud (LMC), Milky Way (MW), Small Magellanic Cloud (SMC) and starburst galaxies is the presence and intensity of an absorption feature at 2175~\AA~rest-wavelength. This feature is generally associated with the absorption of graphite grains, whose abundance and sizes changes between the different models \citep{dra03}. The feature is most prominent from MW to LMC, whereas it is practically absent in SMC \citep{car89} and starburst galaxies \citep{cal01}. SMC and starburst models have a much larger amount of far ultraviolet (FUV) extinction. Previous bursts strongly favoured SMC like dust host galaxies \citep{sch07, sta07, kan08}, and only for very few of them a MW model provides a better fit. Additionally, the host extinction shows a trend of lower extinction at higher redshift \citep{kan06}. However, with increasing redshift, both the FUV absorption as well as the 2175~\AA~bump significantly decrease the detection efficiency for optical follow-up observations, so the present extinction distribution and its dependence on the bursts redshift might be strongly instrumental biased. 

\subsubsection{The SED of the afterglow of GRB~070802}
\label{sed}
In the present analysis we used MW, LMC and SMC like extinction models \citep{sea79, fit86a, fit86b, pre84} as templates to fit our multi-band data. All of the GROND optical and NIR data were obtained simultaneously at the time epoch between 1.5~ks and 3.6~ks post burst. As the generic shape of the early light curve is achromatic, we can exclude large effects from an evolving spectrum. The data were fit by a power law and extinction templates from LMC, SMC, and MW in \textit{HyperZ} \citep{bol00}. The amount of dust, power law slope and normalization were free parameters in the fit. The redshift was fixed to the spectroscopic redshift of 2.45 obtained by the VLT \citep{pro07}. 

The GROND SED is shown in Fig.~\ref{picSpec} and was well fit by the LMC (reduced $\chi^2$=0.67 for 4 d.o.f) and the MW (reduced $\chi^2$=1.96 for 4 d.o.f.) extinction models, while the fit was considerably worse for SMC like extinction (reduced $\chi^2$=4.57 for 4 d.o.f.). The large $\chi^2$ difference between LMC and MW models is mostly due to the g$\arcmin$ band magnitude, where the error in the GROND data is relatively large. The dust extinction in the GRB host in the best fit model is $A_V^{host}$=$0.9\pm0.3$~mag for MW and $A_V^{host}$=$1.8\pm0.3$~mag for LMC models. We caution that these values are derived using local extinction curves for a galaxy at a redshift of 2.45. The extinction curve and thus the amount of dust reddening could be significantly different. Additionally, intervening absorbers could contribute to the observed dust extinction, which is only resolved by the spectrum \citep{fyn07}.

We detect a strong absorption feature in the GROND $i\arcmin$ band. GROND $i\arcmin$ is slightly narrower than the SDSS $i\arcmin$ band and is located at 7630$\pm$537~\AA. The 2175~\AA~bump at redshift 2.45 (i.e. at $\sim$7500~\AA~in the observers frame) provides the ideal and obvious candidate for this broad absorption feature. The 2175~\AA~bump is the dominating spectral signature of dust in the interstellar medium (ISM) in the Milky Way and often attributed to small graphite grains processed by star formation \citep{gor97, dra03, dul04}. However, the nature of the bump is not totally clear. Its strength varies along different sight lines in the Milky Way \citep{car89}. Different size distributions or different chemical compositions could be the origin of this variation \citep{nat84}. 

At high redshift, the search of the 2175~\AA~feature has always been very difficult. It was never clearly detected in single objects, for instance in damped Lyman-$\alpha$ systems (DLAs) along quasar (QSO) sight lines. The presence of the bump was excluded from a composite spectrum of 37 Ca II and Mg II absorbers from Sloan Digital Sky Survey (SDSS) QSOs \citep{wil06}, but detected in a combined spectrum of 18 galaxies at 1$\leq$z$\leq$1.5 with intermediate-age stellar populations \citep{nol07}. Further detections outside the Local Group have been suggested e.g. for GRB~050802 \citep{sch07} at z=1.71 \citep{fyn05} and a galaxy at z=0.83 \citep{mot02}. Despite the presence of strong metal absorption \citep{sav04, ber06, fyn06} and the depletion of refractory elements \citep{sav06} it has not been detected in spectra of previous bursts. The afterglow of GRB~070802 shows the clearest presence of the 2175~\AA~dust feature at a high redshift so far, where it is detected with GROND broad-band photometry and VLT spectroscopy as shown in Fig. 5 in \citet{fyn07} and Eliasdottir et al. (in preperation).

We can estimate the column density of different metals along the sight line for GRB~070802, following the approach described in \citet {sav06}. The $A_{V}^{host}$ is directly proportional to the dust column density. The dust column density is also proportional to the total metal column density. For instance, in a galaxy with MW or LMC like visual extinction, $A_{V}^{host}$=0.5~mag or 0.4~mag is expected for a column density of oxygen of  log($N_{\rm O}$)=17.7 or 17.1, respectively. Assuming a constant dust-to-metals ratio, our $A_{V}^{host}$=0.9~mag for MW or 1.8~mag for LMC gives log($N_{\rm O}$)=18.0 and 17.8. This is the oxygen column density in gas form, i.e.\ assuming that the amount of oxygen locked into dust grains is marginal. In the Galactic ISM, oxygen in dust grains is negligible, but up to about 40\% in the cool gas \citep{sav96}.  If we assume this kind of dust depletion, then the total oxygen column density is log($N_{\rm O}$)=18.4 or 18.2, for MW or LMC, respectively. Using the standard nomenclature for hydrogen absorption at X-ray energies and assuming solar metallicity, we conclude that the column density of hydrogen log($N_{\rm H}$) is about 21.7 or 21.5, for MW or LMC dust, respectively. 

\subsubsection{Combined optical/NIR and X-ray analysis}

The XRT spectrum including all data from 150~s to 405~ks post burst with a total exposure time of 56~ks is compatible with an absorbed power law with a photon index $\Gamma$=2.1 $\pm$ 0.3 (reduced $\chi^2$=0.7 for 15~d.o.f.). The foreground hydrogen column density was frozen at the Galactic value of 3$\times 10^{20}$~cm$^{-2}$ \citep{dic90}. The best fit for $N_{\rm H}$ at z=2.45 is 1.60$^{+1.20}_{-1.10}\times 10^{22}$~cm$^{-2}$, which corresponds to a total gas plus dust oxygen column density of  $N_{\rm O}=7.2^{+5.4}_{-5.0}\times 10^{18}$~cm$^{-2}$. Given the large uncertainties, this is consistent with the values estimated in the previous section.

\begin{figure}
\epsscale{0.8}
\plotone{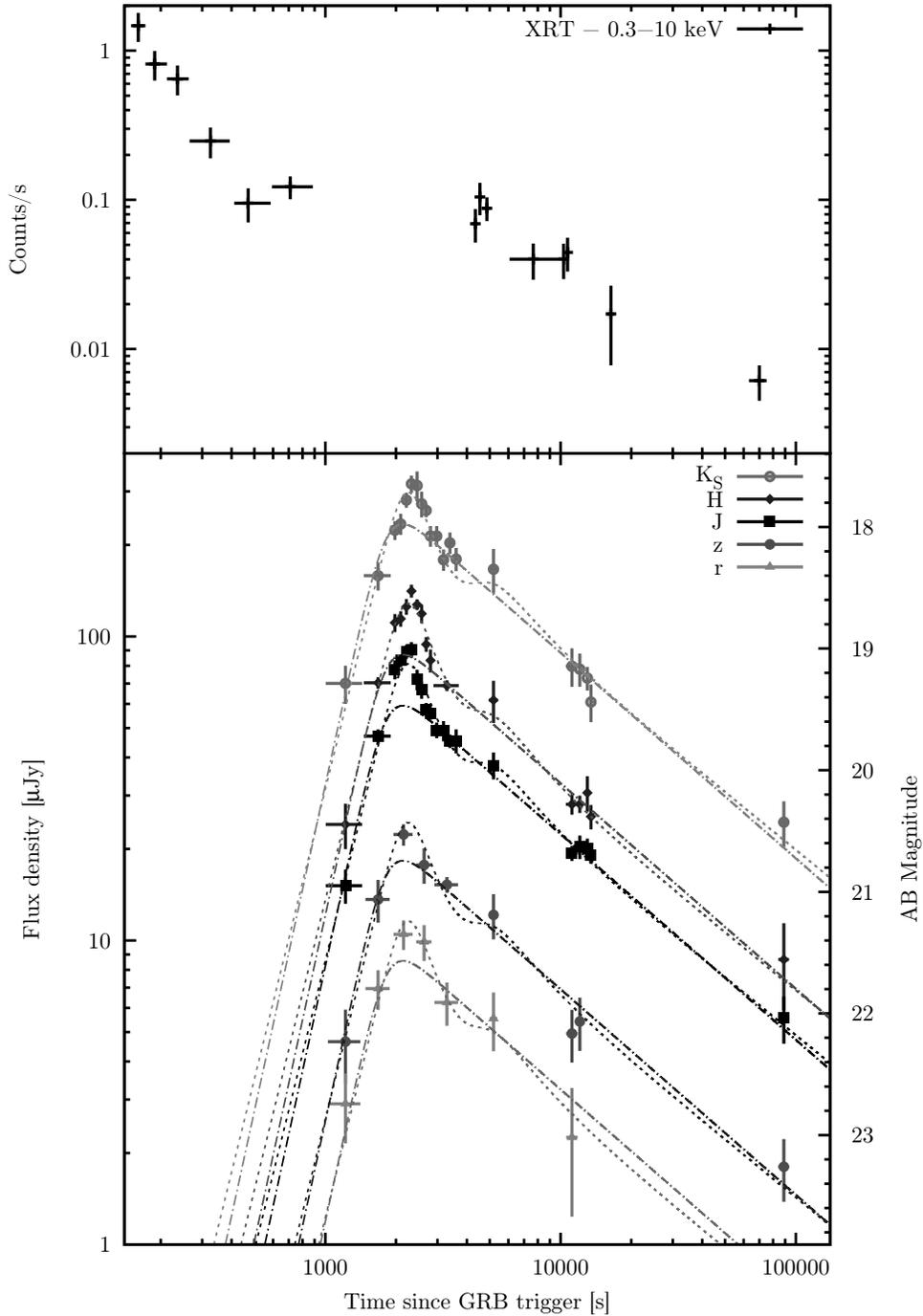}
\caption{X-ray and optical/NIR light curve of the afterglow GRB~070802. The X-ray light curve in the upper panel was obtained from the \textit{Swift} XRT light curve repository \citep{eva07}. The afterglow light curve in the GROND bands $r\arcmin$, $z\arcmin$, $J$, $H$ and $K_S$ is shown in the lower panel. In the $g\arcmin$ and $i\arcmin$ bands the afterglow was too dim to construct a light curve. Also shown are the best fit models in dotted lines for the RS plus FS model and in dashed-dotted lines for the FS model. The simple model of an exclusive FS emission cannot explain the light curve shape around peak brightness and requires an extra emission component superimposed to the overall rise and decay.}
\label{picLight}
\end{figure}

\begin{figure}
\includegraphics[angle=270, width=0.99\columnwidth]{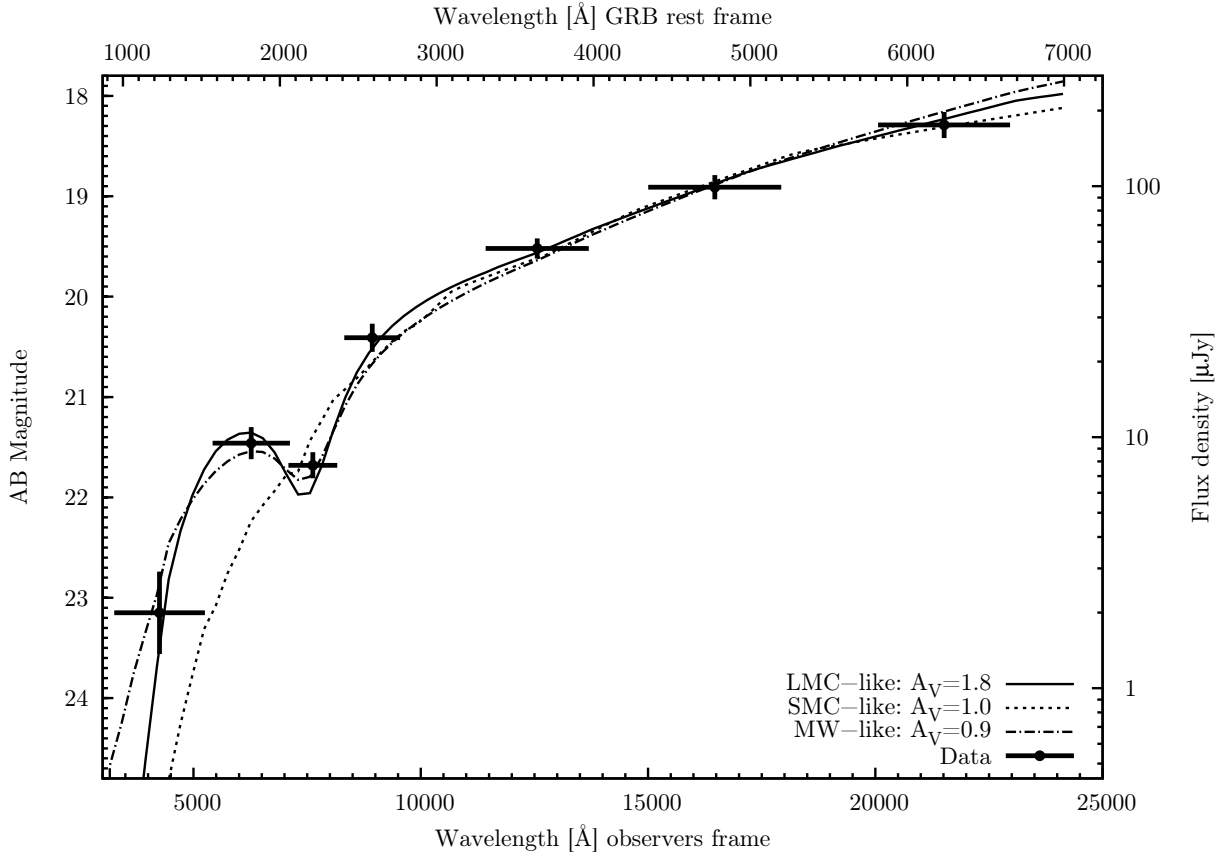}
\caption{Spectral Energy Distribution of the afterglow of GRB~070802 in the GROND filter bands. The data were obtained simultaneously in all colors between T$_0$+1.5~ks and T$_0$+3.6~ks post burst. The shape of the SED can be well reproduced by LMC and MW extinction models. A broad extinction feature is detected in the GROND $i\arcmin$ band at 7630~\AA~central wavelength, which we relate to the 2175~\AA~dust feature in the host at redshift 2.45. Depending on the used model, the best fit extinction varies between $A_V^{host}$ = 0.9~mag (MW) and 1.8~mag (LMC).}
\label{picSpec}
\end{figure}

The GROND optical/NIR and XRT X-ray data can be fit together to constrain the broad-band afterglow spectrum. We fit the data using a power law with absorption in the X-ray regime and extinction in the optical/NIR regime using the XSPEC package \citep{arn96}. Unfortunately, there is no simultaneous early coverage by XRT and GROND, as there is a gap in the XRT coverage of GRB\,070802 from 911\,s to 4.2\,ks (Fig.~\ref{picLight}). Therefore XRT data from 460 s to 5 ks with a total exposure time of 1.3~ks and GROND data from 1.5\,ks to 3.6\,ks after trigger were selected for the joint fit. These intervals were chosen because they are the closest in time while the source is still bright and do not include the steep decay in the early XRT light curve, see Fig.~\ref{picLight} and \citet{imm07}.

The XRT spectrum alone from 460~s to 5~ks post burst is compatible with the total spectrum with a power law photon index of 1.9$\pm{0.4}$, although $N_{\rm H}$ is no longer well constrained due to poorer statistics. 

 The combined GROND and XRT SED from 2.2$~\mu$m $K_S$ to $\sim$4~keV was fit by a single power law with reddening $E_{B-V}$ and $N_{\rm H}$ Galactic foreground values from \citet{sch98} and \citet{dic90} (0.026 and 3$\times 10^{20}$cm$^{-2}$, respectively) and by extinction in the host at z=2.45. There were insufficient statistics above 4~keV in the XRT band to extend the fit to higher energies. The optical/NIR data were modeled using the zdust model, where the extinction in the host is obtained using MW, LMC and SMC reddening laws with values of $R_{V}$ of 3.08, 3.18 and 2.93 respectively, where $A_{V}$ = $E_{B-V}$ x $R_{V}$ \citep{pei92}. As shown in Fig.~\ref{picSpec}, the GROND data alone is well described by MW and LMC models. The additional absorption in the soft X-ray band was modeled by an absorbing column at z=2.45.
  
In the broad-band fits we are mainly interested in the power law index between the GROND and XRT bands. The combined optical/NIR and X-ray data are well described by a single power law with a photon index of 1.91$\pm$0.04 (reduced $\chi^2$ of 1.36 for 12~d.o.f.) using the MW extinction model with $E_{B-V}$ of 0.35$\pm0.04$. The $N_{\rm H}$ at z=2.45 is not well constrained by the fit.  Similar values of the power law index and $E_{B-V}$ are achieved using the LMC and SMC extinction curves. The results of the zdust and single power law models are presented in Table~\ref{xrays} with the MW yielding the best fit.

To test for a cooling break, a broken power law with extinction and soft X-ray absorption was also fit to the data. Here the high energy photon index $\Gamma_2$ was linked to $\Gamma_1$ via $\Gamma_2$=$\Gamma_1$+0.5 as expected in the fireball model.  
The value of the break energy was constrained to lie between the optical and X-ray bands and the best fit photon indices for the low and high energies were $\Gamma_1$=1.61$\pm{0.05}$ and $\Gamma_2$=2.11 respectively. 
The reduced $\chi^2$ for this fit is 1.47 for 11 d.o.f for the MW extinction curve, 2.40 for 11 d.o.f for LMC and 3.5 for 11 d.o.f for SMC. The fit parameters $E_{B-V}$ and the low energy power law index are presented in Table~\ref{xrays}.

The broad-band model without a break provides a better fit to the data, however the difference in the reduced $\chi^2$ parameters of the fits is not conclusive. The single power law implies that the cooling break is redwards the GROND bands at this time, which would be surprisingly early after the explosion. We caution that the GROND and XRT data are not simultaneous. 

We also fit a late time SED using the XRT data 50~ks to 196~ks post burst with the GROND data obtained in the $JHK_S$ band from 86~ks to 96~ks. The best fit model has a power law slope of 2.0$\pm$0.2, so the data are still compatible with the early time single power law model, but we are not able to distinguish between a single and broken power law fit. We do not find evidence for a cooling break between the NIR and X-ray data.

\section{Conclusions}

The optical/NIR afterglow light curve of GRB~070802 can be explained using two models. A combined reverse - forward shock model, and a single forward shock model with a superimposed emission component at peak brightness. Due to the fact that the afterglow peak is heavily delayed compared to the duration of the burst, it is very likely that the increase in brightness in the early light curve is related to the onset of the afterglow as proposed for GRB~060418 and GRB~060607A \citep{mol07}. Using an analogue analysis for GRB~070802, we derive an initial bulk Lorentz factor in the jet of around $\Gamma_0 \approx 160 \left (\frac{E_{53}}{\eta_{0.2}n}\right)^{1/8}$ for an ISM environment. The steep rise of the early light curve favours an homogeneous over a wind shaped circumburst medium. The ground based observations of the optical/NIR afterglow of GRB~070802 were fast enough to detect an early brightening of the afterglow. Further rapid follow-up campaigns may establish whether this rapid rise is a generic feature of GRB afterglow light curves.

A broad-band fit of GROND and XRT data is compatible with a single power law spectrum with photon index 1.91$\pm$0.04, suggesting the cooling break being redwards of the GROND bands at the start of the observations at $\sim$1.2~ks post burst. The late time photon index from NIR to X-rays is still comparable with the early time power law, indicating no time evolution of the spectrum during our observations.

The observed SED from $g\arcmin$~to the $K_S$ band can be well reproduced with LMC and MW extinction models. A broad-band absorption feature in the GROND $i\arcmin$ band with a central wavelength of 7630~\AA~ is unquestionably required to explain the observed SED. The redshifted 2175~\AA~feature in the host galaxy of the burst at z=2.45 known from MW and LMC extinction models provides the ideal candidate. Depending on the model the best fit extinction ranges from $A_V^{host}$=$0.9$~mag for MW like dust absorption to $A_V^{host}$= $1.8$~mag for LMC dust. The amount of extinction is significantly larger than estimated for previous bursts \citep{kan06, sch07, kan08}. Pre-\textit{Swift} bursts have shown a correlation of decreasing extinction with increasing redshift \citep{kan08}. However, there might be a strong instrumental bias, as fast and simultaneous optical to NIR follow-up observations for a large GRB sample is missing. The GROND instrument, with its unique optical and NIR capabilities, is a powerful tool which might remove this bias. GRB~070802 was the first burst for GROND occurring during nighttime and revealed significant amount of dust in its host galaxy, indicating that at least a good fraction of the UVOT dark burst is due to intrinsic extinction in the GRB host galaxy. Future observations will help to quantify the amount of highly extinguished bursts and may help to solve the mystery of dark bursts.

\acknowledgments

This research was supported by the DFG cluster of excellence 'Origin and Structure of the Universe' and by the Hungarian National Office for Research and Technology (NKTH), through the Pol\'anyi Program. We thank the referee for the comments and a very helpful report. SMB acknowledges the support of the European Union through a Marie Curie Intra-European Fellowship within the sixth Framework Program. SK acknowledges financial support by DFG grant Kl 766/13-2. This work made use of data supplied by the UK Swift Science Data Centre at the University of Leicester.




\clearpage

\input{tab1.tex}
\input{tab2.tex}
\input{tab3.tex}
\input{tab4.tex}

\clearpage

\end{document}

%% file: tab1.tex
\begin{deluxetable}{cccccccc}
\tabletypesize{\scriptsize}
\tablecaption{Secondary standards in the GRB field in the GROND filter bands used during the calibration
\label{tabSecStan}}
\tablewidth{0pt}
\tablehead{
\colhead {Identifier} & \colhead {g$\arcmin$} & \colhead {r$\arcmin$} & \colhead {i$\arcmin$} & \colhead {z$\arcmin$} & \colhead{J} & \colhead{H} & \colhead{K$_S$} \\
 &  mag & mag & mag & mag & mag & mag & mag}
\startdata
A & 22.42 $\pm$ 0.05 & 20.71 $\pm$ 0.04 & 19.26 $\pm$ 0.04 & 18.50 $\pm$ 0.04 & 16.98 $\pm$ 0.06 & 16.55 $\pm$ 0.06 & 16.09 $\pm$ 0.08 \\
B & 16.33 $\pm$ 0.07 & 15.89 $\pm$ 0.03 & 15.78 $\pm$ 0.03 & 15.65 $\pm$ 0.03 & 14.81 $\pm$ 0.05 & 14.59 $\pm$ 0.05 & 14.42 $\pm$ 0.06 \\
C & 15.47 $\pm$ 0.05 & 14.98 $\pm$ 0.03 & 14.86 $\pm$ 0.03 & 14.72 $\pm$ 0.03 & 13.87 $\pm$ 0.05 & 13.61 $\pm$ 0.05 & 13.43 $\pm$ 0.06 \\
D & 20.07 $\pm$ 0.06 & 18.76 $\pm$ 0.04 & 18.23 $\pm$ 0.04 & 17.92 $\pm$ 0.04 & 16.72 $\pm$ 0.07 & 16.19 $\pm$ 0.07 & 15.74 $\pm$ 0.09 \\
E & 15.70 $\pm$ 0.05 & 14.62 $\pm$ 0.03 & 14.17 $\pm$ 0.03 & 13.87 $\pm$ 0.03 & 12.71 $\pm$ 0.05 & 12.61 $\pm$ 0.05 & 11.99 $\pm$ 0.06 \\
\hline
\enddata
\end{deluxetable}

%% file: tab2.tex
\begin{deluxetable}{cccccccc}
\tabletypesize{\scriptsize}
\tablecaption{Parameters of the generic light curve fit to the $r\arcmin z\arcmin JHK$ band\label{tab1}}
\tablewidth{0pt}
\tablehead{\colhead{Model} & \colhead{$\alpha^r_1$} & \colhead{$\alpha^r_2$} & \colhead{$\alpha^f_1$} & \colhead{$\alpha^f_2$} & \colhead{$t_1$ [s]} & \colhead{s$^f$}\\
 & RS rise index & RS decline index & FS rise index & FS decline index & & Break sharpness}
\startdata
 Combined RS + FS & 3.16 $\pm$ 0.33 & -2.66 $\pm$ 0.96 & 3.12 $\pm$ 1.49 & -0.63 $\pm$ 0.14 & 2181 $\pm$ 107 & 3.15 $\pm$ 2.16 \\
 Single FS & --- & --- & 3.56 $\pm$ 0.36 & -0.68 $\pm$ 0.04 & 1830 $\pm$ 51.9  &  2.48 $\pm$ 0.72 \\
\hline
\enddata
\end{deluxetable}

%% file: tab3.tex
\begin{deluxetable}{ccccccc}
\tabletypesize{\scriptsize}
\tablecaption{Parameters of the individual fits to the $r\arcmin z\arcmin JHK$ bands.\label{tab2}}
\tablewidth{0pt}
\tablehead{
\colhead {Model} &\colhead {Parameter} &\colhead {$r\arcmin$} & \colhead {$z\arcmin$} & \colhead{$J$} & \colhead{$H$} & \colhead{$K_S$}}
\startdata
Combined RS + FS & $F_r^0$ [$\mu$J] & 14.7 $\pm$ 0.7 & 30.9 $\pm$ 1.6 & 102 $\pm$ 2.5 & 163 $\pm$ 3.1 & 365 $\pm$ 11.0 \\
Combined RS + FS & $F_f^0$ [$\mu$J] & 4.3 $\pm$ 0.7 & 9.7 $\pm$ 1.0 & 33.1 $\pm$ 1.6 & 46.5 $\pm$ 2.1 & 135 $\pm$ 10.9 \\
Combined RS + FS & reduced $\chi^2$/d.o.f. & 0.64/5 & 0.98/7 & 1.80/18 & 1.10/15 & 1.53/18 \\
\hline
 Single FS & $F_f^0$ [$\mu$J] & 10.2 $\pm$ 0.6 & 21.8 $\pm$ 0.9 & 70.6 $\pm$ 1.0 & 104 $\pm$ 2.0 & 278 $\pm$ 7.6 \\
 Single FS & reduced $\chi^2$/d.o.f. & 0.62/3 & 0.66/5 & 0.98/7 & 0.67/7 & 0.80/7 \\
\hline
\enddata
\end{deluxetable}

%% file: tab4.tex
\begin{deluxetable}{cccccccc}
\tabletypesize{\scriptsize}
\tablecaption{Broad band spectral fits to the GROND and XRT data using XSPEC. 
\label{xrays}}
\tablewidth{0pt}
\tablehead{
\colhead {Extinction Model} & Power law & \colhead {E(B-V)} & \colhead {Photon Index} & \colhead { $\chi^2$/d.o.f}  \\
 }
\startdata

 MW  & 	Single & 0.35$\pm{0.04}$ & 1.91$\pm{0.04}$ & 1.36/12 \\
MW  & 	Broken & 0.41$\pm{0.04}$ & 1.61$\pm{0.05}$ & 1.47/11 \\
\hline
LMC  & 	Single & 0.35$\pm{0.04}$ & 1.92$\pm{0.03}$ & 2.13/12 \\
 LMC  & 	Broken & 0.39$\pm{0.04}$ & 1.61$\pm{0.05}$ &  2.40/11 \\
\hline
 SMC  & 	Single & 0.34$^{+0.04}_{-0.03}$ & 1.90$^{+0.03}_{-0.04}$ & 3.2/12 \\
 SMC  & 	Broken & 0.34$^{+0.04}_{-0.02}$ & 1.61$\pm{0.03}$ &  3.5/11 \\

\hline

\enddata
\end{deluxetable}

%% file: ms.bbl
\begin{thebibliography}{81}
\expandafter\ifx\csname natexlab\endcsname\relax\def\natexlab#1{#1}\fi

\bibitem[{{Arnaud}(1996)}]{arn96}
{Arnaud}, K.~A. 1996, in Astronomical Society of the Pacific Conference Series,
  Vol. 101, Astronomical Data Analysis Software and Systems V, ed. G.~H.
  {Jacoby} \& J.~{Barnes}, 17

\bibitem[{{Barthelmy} {et~al.}(2005)}]{bar05}
{Barthelmy}, S.~D. {et~al.} 2005, Space Science Reviews, 120, 143

\bibitem[{{Barthelmy} {et~al.}(2007)}]{bar07}
---. 2007, GRB Coordinates Network, 6692

\bibitem[{{Beloborodov}(2005)}]{bel05}
{Beloborodov}, A.~M. 2005, \apjl, 618, L13

\bibitem[{{Berger} \& {Murphy}(2007)}]{ber07}
{Berger}, E. \& {Murphy}, D. 2007, GRB Coordinates Network, 6695

\bibitem[{{Berger} {et~al.}(2006){Berger}, {Penprase}, {Cenko}, {Kulkarni},
  {Fox}, {Steidel}, \& {Reddy}}]{ber06}
{Berger}, E., {Penprase}, B.~E., {Cenko}, S.~B., {Kulkarni}, S.~R., {Fox},
  D.~B., {Steidel}, C.~C., \& {Reddy}, N.~A. 2006, \apj, 642, 979

\bibitem[{{Beuermann} {et~al.}(1999)}]{beu99}
{Beuermann}, K. {et~al.} 1999, \aap, 352, L26

\bibitem[{{Bolzonella} {et~al.}(2000){Bolzonella}, {Miralles}, \&
  {Pell{\'o}}}]{bol00}
{Bolzonella}, M., {Miralles}, J.-M., \& {Pell{\'o}}, R. 2000, \aap, 363, 476

\bibitem[{{Bromm} \& {Loeb}(2002)}]{bro02}
{Bromm}, V. \& {Loeb}, A. 2002, \apj, 575, 111

\bibitem[{{Burrows} {et~al.}(2005)}]{bur05a}
{Burrows}, D.~N. {et~al.} 2005, Space Science Reviews, 120, 165

\bibitem[{{Calzetti}(2001)}]{cal01}
{Calzetti}, D. 2001, \pasp, 113, 1449

\bibitem[{{Cardelli} {et~al.}(1989){Cardelli}, {Clayton}, \& {Mathis}}]{car89}
{Cardelli}, J.~A., {Clayton}, G.~C., \& {Mathis}, J.~S. 1989, \apj, 345, 245

\bibitem[{{Cummings} {et~al.}(2007)}]{cum07}
{Cummings}, J. {et~al.} 2007, GRB Coordinates Network, 6699

\bibitem[{{de Ugarte Postigo} {et~al.}(2005)}]{uga05}
{de Ugarte Postigo}, A. {et~al.} 2005, \aap, 443, 841

\bibitem[{{Dickey} \& {Lockman}(1990)}]{dic90}
{Dickey}, J.~M. \& {Lockman}, F.~J. 1990, \araa, 28, 215

\bibitem[{{Draine}(2003)}]{dra03}
{Draine}, B.~T. 2003, \araa, 41, 241

\bibitem[{{Duley} \& {Lazarev}(2004)}]{dul04}
{Duley}, W.~W. \& {Lazarev}, S. 2004, \apjl, 612, L33

\bibitem[{{Evans} {et~al.}(2007)}]{eva07}
{Evans}, P.~A. {et~al.} 2007, \aap, 469, 379

\bibitem[{{Ferrero} {et~al.}(2008)}]{fer08}
{Ferrero}, P. {et~al.} 2008, \aap, submitted, (arXiv:0804.2457)

\bibitem[{{Fishman}(1994)}]{fis94}
{Fishman}, G. {et~al.} 1986, \apjs, 92, 229

\bibitem[{{Fitzpatrick}(1986)}]{fit86b}
{Fitzpatrick}, E.~L. 1986, \aj, 92, 1068

\bibitem[{{Fitzpatrick} \& {Massa}(1986)}]{fit86a}
{Fitzpatrick}, E.~L. \& {Massa}, D. 1986, \apj, 307, 286

\bibitem[{{Fox} {et~al.}(2003)}]{fox03}
{Fox}, D.~B. {et~al.} 2003, \nat, 422, 284

\bibitem[{{Fynbo} {et~al.}(2001)}]{fyn01}
{Fynbo}, J.~P.~U. {et~al.} 2001, \aap, 373, 796

\bibitem[{{Fynbo} {et~al.}(2005)}]{fyn05}
---. 2005, GRB Coordinates Network, 3749

\bibitem[{{Fynbo} {et~al.}(2006)}]{fyn06}
---. 2006, \aap, 451, L47

\bibitem[{{Fynbo} {et~al.}(2007)}]{fyn07}
---. 2007, ESO Messenger, 130, 43

\bibitem[{{Galama} \& {Wijers}(2001)}]{gal01}
{Galama}, T.~J. \& {Wijers}, R.~A.~M.~J. 2001, \apjl, 549, L209

\bibitem[{{Gehrels} {et~al.}(2004)}]{geh04}
{Gehrels}, N. {et~al.} 2004, \apj, 611, 1005

\bibitem[{{Gordon} {et~al.}(1997){Gordon}, {Calzetti}, \& {Witt}}]{gor97}
{Gordon}, K.~D., {Calzetti}, D., \& {Witt}, A.~N. 1997, \apj, 487, 625

\bibitem[{{Greiner} {et~al.}(2007{\natexlab{a}}){Greiner}, {Clemens},
  {Kr\"{u}hler}, {K\"{u}pc\"{u} Yolda\c{s}}, {Primak}, {Szokoly}, {Yolda\c{s}},
  \& {Klose}}]{gre07}
{Greiner}, J., {Clemens}, C., {Kr\"{u}hler}, T., {K\"{u}pc\"{u} Yolda\c{s}},
  A., {Primak}, N., {Szokoly}, G., {Yolda\c{s}}, A., \& {Klose}, S.
  2007{\natexlab{a}}, GRB Coordinates Network, 6694

\bibitem[{{Greiner} {et~al.}(2007{\natexlab{b}})}]{gre07a}
{Greiner}, J. {et~al.} 2007{\natexlab{b}}, ESO Messenger, 130, 12

\bibitem[{{Greiner} {et~al.}(2008)}]{gre08}
---. 2008, \pasp, 120, 405

\bibitem[{{Groot} {et~al.}(1998)}]{gro98}
{Groot}, P.~J. {et~al.} 1998, \apjl, 493, L27

\bibitem[{{Guidorzi} {et~al.}(2005)}]{gui05}
{Guidorzi}, C. {et~al.} 2005, \apjl, 630, L121

\bibitem[{{Immler} {et~al.}(2007){Immler}, {Mangano}, {Kuin}, \&
  {Cummings}}]{imm07}
{Immler}, S., {Mangano}, V., {Kuin}, N.~P.~M., \& {Cummings}, J. 2007, GCN
  Report, 78

\bibitem[{{Kann} {et~al.}(2006){Kann}, {Klose}, \& {Zeh}}]{kan06}
{Kann}, D.~A., {Klose}, S., \& {Zeh}, A. 2006, \apj, 641, 993

\bibitem[{{Kann} {et~al.}(2008)}]{kan08}
{Kann}, D.~A. {et~al.} 2008, \apj ,~submitted, (arXiv:0804.1959)

\bibitem[{{Katz}(1994)}]{kat94}
{Katz}, J.~I. 1994, \apjl, 432, L107

\bibitem[{{Kawai} {et~al.}(2006)}]{kaw06}
{Kawai}, N. {et~al.} 2006, \nat, 440, 184

\bibitem[{{Klose} {et~al.}(2003)}]{klo03}
{Klose}, S. {et~al.} 2003, \apj, 592, 1025

\bibitem[{{Kobayashi} \& {Sari}(2000)}]{kob00}
{Kobayashi}, S. \& {Sari}, R. 2000, \apj, 542, 819

\bibitem[{{Kouveliotou} {et~al.}(1993){Kouveliotou}, {Meegan}, {Fishman},
  {Bhat}, {Briggs}, {Koshut}, {Paciesas}, \& {Pendleton}}]{kou93}
{Kouveliotou}, C., {Meegan}, C.~A., {Fishman}, G.~J., {Bhat}, N.~P., {Briggs},
  M.~S., {Koshut}, T.~M., {Paciesas}, W.~S., \& {Pendleton}, G.~N. 1993, \apjl,
  413, L101

\bibitem[{{Kumar} \& {Piran}(2000)}]{kum00}
{Kumar}, P. \& {Piran}, T. 2000, \apj, 535, 152

\bibitem[{{Lamb} \& {Reichart}(2000)}]{lam00}
{Lamb}, D.~Q. \& {Reichart}, D.~E. 2000, \apj, 536, 1

\bibitem[{{Lazzati} {et~al.}(2002){Lazzati}, {Rossi}, {Covino}, {Ghisellini},
  \& {Malesani}}]{laz02}
{Lazzati}, D., {Rossi}, E., {Covino}, S., {Ghisellini}, G., \& {Malesani}, D.
  2002, \aap, 396, L5

\bibitem[{{Lipkin} {et~al.}(2004)}]{lip04}
{Lipkin}, Y.~M. {et~al.} 2004, \apj, 606, 381

\bibitem[{{MacFadyen} \& {Woosley}(1999)}]{mac99}
{MacFadyen}, A.~I. \& {Woosley}, S.~E. 1999, \apj, 524, 262

\bibitem[{{Mangano} {et~al.}(2007){Mangano}, {Sbarufatti}, {La Parola},
  {Troja}, {Evans}, \& {Immler}}]{man07}
{Mangano}, V., {Sbarufatti}, B., {La Parola}, V., {Troja}, E., {Evans}, P., \&
  {Immler}, S. 2007, GRB Coordinates Network, 6702

\bibitem[{{M{\'e}sz{\'a}ros}(2006)}]{mes06}
{M{\'e}sz{\'a}ros}, P. 2006, Rep. Prog. Phys., 69, 2259

\bibitem[{{M{\'e}sz{\'a}ros} \& {Rees}(1997)}]{mes97}
{M{\'e}sz{\'a}ros}, P. \& {Rees}, M.~J. 1997, \apj, 476, 232

\bibitem[{{Molinari} {et~al.}(2007)}]{mol07}
{Molinari}, E. {et~al.} 2007, \aap, 469, L13

\bibitem[{{Motta} {et~al.}(2002)}]{mot02}
{Motta}, V. {et~al.} 2002, \apj, 574, 719

\bibitem[{{Nakar} \& {Granot}(2007)}]{nak07}
{Nakar}, E. \& {Granot}, J. 2007, \mnras, 380, 1744

\bibitem[{{Nakar} \& {Piran}(2004)}]{nak04}
{Nakar}, E. \& {Piran}, T. 2004, \mnras, 353, 647

\bibitem[{{Natta} \& {Panagia}(1984)}]{nat84}
{Natta}, A. \& {Panagia}, N. 1984, \apj, 287, 228

\bibitem[{{Noll} {et~al.}(2007){Noll}, {Pierini}, {Pannella}, \&
  {Savaglio}}]{nol07}
{Noll}, S., {Pierini}, D., {Pannella}, M., \& {Savaglio}, S. 2007, in
  ASP Conf. Ser., 380, Deepest
  Astronomical Surveys, ed. J.~{Afonso}, H.~C. {Ferguson}, B.~{Mobasher}, \&
  R.~{Norris}, 461

\bibitem[{{O'Brien} {et~al.}(2006)}]{bri06}
{O'Brien}, P.~T. {et~al.} 2006, \apj, 647, 1213

\bibitem[{{Paczynski}(1998)}]{pac98}
{Paczynski}, B. 1998, \apjl, 494, L45

\bibitem[{{Panaitescu} \& {Kumar}(2000)}]{pan00}
{Panaitescu}, A. \& {Kumar}, P. 2000, \apj, 543, 66

\bibitem[{{Pe'er} {et~al.}(2007){Pe'er}, {Ryde}, {Wijers}, {M{\'e}sz{\'a}ros},
  \& {Rees}}]{pee07}
{Pe'er}, A., {Ryde}, F., {Wijers}, R.~A.~M.~J., {M{\'e}sz{\'a}ros}, P., \&
  {Rees}, M.~J. 2007, \apjl, 664, L1

\bibitem[{{Perley} {et~al.}(2008)}]{per08}
{Perley}, D.~A. {et~al.} 2008, \apj, 672, 449

\bibitem[{{Pei}(1992)}]{pei92}
{Pei}, Y.~C. 1992, \apj, 395, 130

\bibitem[{{Piran}(2000)}]{pir00}
{Piran}, T. 2000, \physrep, 333, 529

\bibitem[{{Piran}(2005)}]{pir05}
{Piran}, T. 2005, Rev. Mod. Phys., 76, 1143

\bibitem[{{Prevot}(1984)}]{pre84}
{Prevot}, M.~L. and {Lequeux}, J. and {Prevot}, L. and {Maurice}, E. and 
	{Rocca-Volmerange}, B. 1984, \aap, 132, 389

\bibitem[{{Prochaska} {et~al.}(2007){Prochaska}, {Th\"{o}ne}, {Malesani},
  {Fynbo}, \& {Vreeswijk}}]{pro07}
{Prochaska}, J.~X., {Th\"{o}ne}, C.~C., {Malesani}, D., {Fynbo}, J.~P.~U., \&
  {Vreeswijk}, P.~M. 2007, GRB Coordinates Network, 6698

\bibitem[{{Rees} \& {M{\'e}sz{\'a}ros}(1998)}]{ree98}
{Rees}, M.~J. \& {M{\'e}sz{\'a}ros}, P. 1998, \apjl, 496, L1

\bibitem[{{Rol} {et~al.}(2005){Rol}, {Wijers}, {Kouveliotou}, {Kaper}, \&
  {Kaneko}}]{rol05}
{Rol}, E., {Wijers}, R.~A.~M.~J., {Kouveliotou}, C., {Kaper}, L., \& {Kaneko},
  Y. 2005, \apj, 624, 868

\bibitem[{{Roming} {et~al.}(2005)}]{rom05}
{Roming}, P.~W.~A. {et~al.} 2005, Space Science Reviews, 120, 95

\bibitem[{{Roming} {et~al.}(2006)}]{rom06}
---. 2006, \apj, 652, 1416


\bibitem[{{Sari} \& {Piran}(1999)}]{sar99}
{Sari}, R. \& {Piran}, T. 1999, \apj, 520, 641

\bibitem[{{Sari} {et~al.}(1998){Sari}, {Piran}, \& {Narayan}}]{sar98}
{Sari}, R., {Piran}, T., \& {Narayan}, R. 1998, \apjl, 497, L17

\bibitem[{{Savage} \& {Sembach}(1996)}]{sav96}
{Savage}, B.~D. \& {Sembach}, K.~R. 1996, \araa, 34, 279

\bibitem[{{Savaglio}(2006)}]{sav06}
{Savaglio}, S. 2006, New Journal of Physics, 8, 195

\bibitem[{{Savaglio} \& {Fall}(2004)}]{sav04}
{Savaglio}, S. \& {Fall}, S.~M. 2004, \apj, 614, 293

\bibitem[{{Schady} {et~al.}(2007)}]{sch07}
{Schady}, P. {et~al.} 2007, \mnras, 377, 273

\bibitem[{{Schlegel} {et~al.}(1998){Schlegel}, {Finkbeiner}, \&
  {Davis}}]{sch98}
{Schlegel}, D.~J., {Finkbeiner}, D.~P., \& {Davis}, M. 1998, \apj, 500, 525

\bibitem[{{Seaton}(1979)}]{sea79}
{Seaton}, M.~J. 1979, \mnras, 187, 73P

\bibitem[{{Smith} {et~al.}(2002)}]{smi02}
{Smith}, J.~A. {et~al.} 2002, \aj, 123, 2121

\bibitem[{{Starling} {et~al.}(2007){Starling}, {Wijers}, {Wiersema}, {Rol},
  {Curran}, {Kouveliotou}, {van der Horst}, \& {Heemskerk}}]{sta07}
{Starling}, R.~L.~C., {Wijers}, R.~A.~M.~J., {Wiersema}, K., {Rol}, E.,
  {Curran}, P.~A., {Kouveliotou}, C., {van der Horst}, A.~J., \& {Heemskerk},
  M.~H.~M. 2007, \apj, 661, 787

\bibitem[{{Stratta} {et~al.}(2004){Stratta}, {Fiore}, {Antonelli}, {Piro}, \&
  {De Pasquale}}]{str04}
{Stratta}, G., {Fiore}, F., {Antonelli}, L.~A., {Piro}, L., \& {De Pasquale},
  M. 2004, \apj, 608, 846

\bibitem[{{Tody}(1993)}]{tod93}
{Tody}, D. 1993, in ASP Conf. Ser., 52, Astronomical Data Analysis Software and Systems II, ed. R.~J.
  {Hanisch}, R.~J.~V. {Brissenden}, \& J.~{Barnes}, 173

\bibitem[{{Updike} {et~al.}(2008)}]{upd08}
{Updike}, A.~C. {et~al.} 2008, \apj, accepted, (arXiv:0805.1094)

\bibitem[{{van Paradijs} {et~al.}(2000){van Paradijs}, {Kouveliotou}, \&
  {Wijers}}]{van00}
{van Paradijs}, J., {Kouveliotou}, C., \& {Wijers}, R.~A.~M.~J. 2000, \araa,
  38, 379

\bibitem[{{Wang} \& {Loeb}(2000)}]{wan00}
{Wang}, X. \& {Loeb}, A. 2000, \apj, 535, 778


\bibitem[{{Wijers} {et~al.}(1997){Wijers}, {Rees}, \& {M{\'e}sz{\'a}ros}}]{wij97}
{Wijers}, R.~A.~M.~J., {Rees}, M.~J., \& {M{\'e}sz{\'a}ros}, P. 1997, \mnras, 288, L51

\bibitem[{{Wild} {et~al.}(2006){Wild}, {Hewett}, \& {Pettini}}]{wil06}
{Wild}, V., {Hewett}, P.~C., \& {Pettini}, M. 2006, \mnras, 367, 211

\bibitem[{{Woosley} \& {Bloom}(2006)}]{woo06}
{Woosley}, S.~E. \& {Bloom}, J.~S. 2006, \araa, 44, 507

\bibitem[{{Zeh} {et~al.}(2004)}]{zeh04}
{Zeh}, A., {Klose}, S., \& {Hartmann}, D.~H. 2004, \apj, 609, 952

\bibitem[{{Zhang}(2007)}]{zha07}
{Zhang}, B. 2007, Chinese Journal of Astronomy and Astrophysics, 7, 1

\bibitem[{{Zhang} {et~al.}(2006){Zhang}, {Fan}, {Dyks}, {Kobayashi},
  {M{\'e}sz{\'a}ros}, {Burrows}, {Nousek}, \& {Gehrels}}]{zha06}
{Zhang}, B., {Fan}, Y.~Z., {Dyks}, J., {Kobayashi}, S., {M{\'e}sz{\'a}ros}, P.,
  {Burrows}, D.~N., {Nousek}, J.~A., \& {Gehrels}, N. 2006, \apj, 642, 354

\bibitem[{{Zhang} {et~al.}(2003){Zhang}, {Kobayashi}, \&
  {M{\'e}sz{\'a}ros}}]{zha03}
{Zhang}, B., {Kobayashi}, S., \& {M{\'e}sz{\'a}ros}, P. 2003, \apj, 595, 950



\end{thebibliography}
